\documentclass[preprint,amsmath,amssymb,dvips]{revtex4}
\usepackage{graphicx}
\usepackage{bm}
\begin{document}
\title{Magnetic field dependence of the entanglement entropy of one dimensional spin systems in quantum phase transition induced by a quench}
\author{Banasri Basu$^\dag$}
\affiliation{Physics and Applied Mathematics Unit, Indian Statistical Institute, Kolkata 700 108, India.}
\author{Pratul Bandyopadhyay$^\star$}
\affiliation{Physics and Applied Mathematics Unit, Indian Statistical Institute, Kolkata 700 108, India.}
\author{Priyadarshi Majumdar$^\prime$}
\affiliation{Jyotinagar Bidyasree Niketan H.S. School, 41 Jyotinagar, Kolkata 700 108, India.}

\vskip .5cm

\begin{abstract}

\noindent We study the magnetic field dependence of the entanglement entropy in quantum phase transition induced by a quench of the XX, XXX and the LMG model. The entropy for a block of $L$ spins with the rest follows a logarithmic scaling law where the block size $L$ is restricted due to the dependence of the prefactor on the quench time. Within this restricted region the entropy undergoes a renormalization group (RG) flow. From the RG flow equation we have analytically determined the magnetic field dependence of the entropy. The anisotropy parameter dependence of the entropy for the XY and the LMG model has also been studied in this framework. The results are found to be in excellent agreement with that obtained by other authors from numerical studies without any quench.

keywords: entanglement entropy, renormalization group, magnetic field;

Pacs nos: 03.67.Bg,~03.67.Hk,~03.65.Ud,~73.43.Nq;

$\dag$: E-mail : banasri@isical.ac.in;
$\star$: E-mail: b$\_$ pratul@yahoo.co.in

$\prime$: E-mail: majumdar$\_$ priyadarshi@yahoo.com

\end{abstract}

\maketitle

\newpage

\section{Introduction:} \label{sec_1}

\noindent It is well known that the entanglement entropy in quantum phase transition (QPT) of one dimensional spin systems in general follows a scaling law. In fact the entanglement entropy of a block of $L$ spins with the rest of the system follows a logarithmic scaling law where the prefactor is determined by the central charge of the relevant conformal field theory. Critical ground states are characterized by an entropy $S_L$ that diverges logarithmically with $L$ having the relation with a coefficient given by the holomorphic and antiholomorphic central charge of the conformal field theory~\cite{1}. Indeed this expression corresponds to the geometric entropy for a conformal field theory derived by Holzhey, Larsen and Wilczek~\cite{2}. It has been shown in some earlier papers~\cite{3,4} that the central charge in conformal field theory corresponds to the Berry phase factor acquired by a spin state when it evolves in a closed path. The entanglement entropy of a pure state can be reduced to the measure of entanglement of formation in a mixed state given by concurrence~\cite{5}. In a spin system the concurrence $C$ for the entanglement of two nearest neighbor spins is found to be given by the Berry phase factor $\widetilde{\phi}$ where the Berry phase acquired by a spin state when it evolves in a closed path is $e^{i\phi}=e^{i2\pi \widetilde{\phi}}$~\cite{6,7,8}. It may be noted that the central charge in conformal field theory satisfies a renormalization group (RG) flow as pointed out by Zamolodchikov~\cite{9}. This implies that the Berry phase factor $\widetilde{\phi}$ as well as the concurrence for an entangled spin system also satisfies the RG flow equation. This essentially corresponds to the fact that the entanglement entropy undergoes a RG flow~\cite{10}.
\\
\noindent In some recent works~\cite{11,12} it has been pointed out that the entanglement entropy in QPT in one dimensional spin systems induced by a quench also satisfies a scaling law in a restricted sense such that there is a constraint on the block size $L$ depending on the quench time. Indeed in this case the prefactor has a dependence on the quench time. An interesting result observed in this case is that the entanglement entropy in QPT in all one dimensional spin systems induced by a quench follows a universal behavior. In the scaling region the entanglement entropy undergoes a RG flow.
\\
\noindent It may be noted that the external magnetic field is the control parameter in QPT. In transverse Ising model we have a sharp critical point when the external field parameter $\lambda$ takes the value $\lambda=\lambda_C=1$. For $\lambda >1$ the system is in a paramagnetic state and for $\lambda <1$ the system transits to a ferromagnetic state having all the spins either in the up or down direction. However for the XX model and the Heisenberg spin chain (XXX model) we have a critical region where the system is gapless. In fact in this case $0<\lambda<2$ corresponds to the critical region. For $\lambda=0$ the system attains the maximum entropy and as $\lambda$ increases the entropy decreases when at $\lambda=2$ it vanishes. In fact at $\lambda=2$ both the XX and XXX systems transit to ferromagnetic states. An analogous behavior appears in the Lipkin-Meshkov-Glick (LMG) model~\cite{13} when in the isotropic case criticality corresponds to the region $0<\lambda<1$. Here also in the isotropic case the entropy is maximum at $\lambda=0$ and as $\lambda$ increases entropy decreases and finally it vanishes at $\lambda=1$. For $\lambda \geq 1$ the ground state is a fully polarized product state.
\\
\noindent In some recent works~\cite{11,12} we have computed the entanglement entropy in QPT induced by a quench for the XX, XXX and LMG models at $\lambda=0$. From an analysis of the RG flow we shall study here the behavior of the entropy with the increase in $\lambda$ until it vanishes. Also we shall extend our study for the XY and LMG models by quenching the system across quantum multicritical points by approaching along a linear path. In this case the external field is a linear function of the anisotropy parameter $\gamma$. We shall study the behavior of the entropy with the change in the anisotropy parameter.
\\
\noindent In sec.\ref{sec_2} we shall consider the behavior of the entropy with the change in the external field ($\lambda$) for the XX and XXX models. In sec.\ref{sec_3} we shall consider the LMG model. In sec.\ref{sec_4} we shall study the behavior of the entropy at criticality in the XY model as well as in the LMG model with the change of the anisotropy parameter.

\vskip .5cm

\section{Entanglement entropy in the critical region of XX and XXX models:}  \label{sec_2}

\noindent The XX model is given by the Hamiltonian
\begin{equation}
H=-\sum_i \left(\sigma_i^x\sigma_{i+1}^x+\sigma_i^y\sigma_{i+1}^y\right)+\lambda~\sum_i\sigma_i^z. \label{eq_1}
\end{equation}
The criticality of the system has two-limit behavior. At $\lambda=2$ the system corresponds to a ferromagnetic state while at $\lambda=0$ the system falls into the free boson universality class. The interval between these two points corresponds to the critical region. The entropy is maximum at $\lambda=0$ and with the increase of the magnetic field the entropy decreases and finally at $\lambda=2$ it vanishes. At $\lambda=2$ the system transits to the ferromagnetic state when the ground state corresponds to the product state. At $\lambda=0$ the entropy scales like
\begin{equation}
S_L \sim \frac{c+\bar{c}}{6}log_2L \label{eq_2}
\end{equation}
which corresponds to the entropy of a block of $L$ spins with the rest of the system. Here $c~(\bar{c})$ is the holomorphic (antiholomorphic) central charge of the relevant conformal field theory and in the bosonic case we have $c=\bar{c}=1$. Thus we have
\begin{equation}
S_L \sim \frac{1}{3}log_2L. \label{eq_3}
\end{equation}
It has been shown in earlier papers~\cite{3,4} that the central charge in conformal field theory is related to the Berry phase factor $\widetilde{\phi}$, the phase being $e^{i2\pi \widetilde{\phi}}$ which is acquired by a spin state when it evolves in a closed path. It may be pointed out that the measure of entanglement given by concurrence of nearest neighbor spins in a mixed state is equivalent to the entanglement entropy in a pure state~\cite{5}. It has been shown in some earlier papers~\cite{6,7,8} that the concurrence $C$ corresponding to the entanglement of two nearest neighbor spins is related to the Berry phase factor $\widetilde{\phi}$ and we have $C=\widetilde{\phi}$. Now we note that in view of the relation of the central charge $c$ with the Berry phase factor $\widetilde{\phi}$ Zamolodchikov's $c$-theorem~\cite{9} representing the RG flow of the central charge can be transcribed in terms of $\widetilde{\phi}$ and the concurrence $C$ which essentially implies the RG flow of the entanglement entropy~\cite{10}. The RG flow suggests that the entropy decreases along the flow and we have
\begin{equation}
L\frac{\partial \widetilde{\phi}}{\partial L} \leq 0, \label{eq_4}
\end{equation}
where $L$ is a length scale. From this we have
\begin{equation}
\left|\widetilde{\phi}\right|_L \approx a~ln~L=\bar{a}log_2L. \label{eq_5}
\end{equation}
For a pair of nearest neighbor spins ($L=2$) we have $\bar{a}=\left|\widetilde{\phi}\right|$ which corresponds to the concurrence of two nearest neighbor spins in the system. So for the entanglement entropy for a block of $L$ spins with the rest we write
\begin{equation}
S_L \approx \left|\widetilde{\phi}\right|log_2L. \label{eq_6}
\end{equation}
At the critical point $\lambda=0$, the system belongs to the boson universality class and the Berry phase factor $\left|\widetilde{\phi}\right|$ which is identical with the concurrence for a pair of nearest neighbor spins in an antiferromagnetic system is given by $C=\left|\widetilde{\phi}\right|=0.386$~\cite{14,15}. This is very close to the prefactor 1/3 in \eqref{eq_3} derived from the conformal field theory. It may be mentioned here that the entanglement of a block of $L$ spins with the rest of the system can be considered to be equivalent to the entanglement between a single spin representing the block spin with another spin represented by the rest of the system in block variable RG scheme. In view of this $S_L$ in \eqref{eq_6} can be considered as the concurrence $C$ for the entanglement between the pair of this two block variable renormalized spins in a mixed state. The slight departure of the prefactor 1/3 in \eqref{eq_3} from the value $\left|\widetilde{\phi}\right|=0.386$ in \eqref{eq_6} may be associated with the block variable renormalization of the spin sytem which induces change in the coupling constant. In fact in our earlier works~\cite{11,12} from an analysis of the transverse Ising model we have introduced a correction factor 0.926 associated with the block spin variable.
\\
\noindent The introduction of a quench incorporates a new length scale given by the Kibble-Zurek (KZ) correlation length $\widehat{\xi}$~\cite{16,17,18,19,20}, which scales like $\widehat{\xi} \sim \sqrt{\tau}$, $\tau$ being the quench time~\cite{12}. Taking into account this aspect we have found the entanglement entropy for a block of $L$ spins with the rest at $\lambda=0$ for the XX model~\cite{12}
\begin{equation}
S_L~ (\lambda=0)=\frac{2\left|\widetilde{\phi}\right|log_2L}{\left|\widetilde{\phi}\right|log_2\widehat{\xi}}\times 0.926 \approx 3.7\frac{log_2L}{log_2\tau}. \label{eq_7}
\end{equation}
It should be mentioned that the value of $L$ is here restricted and therefore the entropy does not rise with the addition of spin in the block size indefinitely. In fact from the constraint $S_L(\tau)/S_{max}$ where the maximum value of the entropy is given by~\cite{12}
\begin{equation}
S_{max}=2(|\widetilde{\phi}|log_2\widehat{\xi}+1)\times 0.926\approx 0.25 ~ln\tau+1.85, \label{eq_8}
\end{equation}
we have the relation
\begin{equation}
ln~L \leq 0.07(ln~\tau)^2+0.5ln~\tau.  \label{eq_9}
\end{equation}
Within this restricted region the entropy undergoes the RG flow.
\\
\noindent Now to study the magnetic field dependence of the entropy in the critical region $0<\lambda<2$ the time dependent magnetic field is taken to be given by
\begin{equation}
\lambda(t<0)=2-\frac{2t}{\tau}, \label{eq_10}
\end{equation}
so that at $t=\tau$ the system resides at the critical point $\lambda=0$ and it evolves toward $\lambda=2$ when at the end $t=0$ it reaches there. Now transcribing the RG flow equation \eqref{eq_4} in terms of time ($L=ct$), we write
\begin{equation}
\left(\frac{t}{\tau}\right)\frac{\partial \widetilde{\phi}}{\partial \left(t/\tau \right)} \leq 0, \label{eq_11}
\end{equation}
which implies
\begin{equation}
\left|\widetilde{\phi}\right|_t \approx a ~ln~t/\tau=\bar{a}log_2~t/\tau. \label{eq_12}
\end{equation}
As before identifying $\bar{a}=\left|\widetilde{\phi}\right|$ we have
\begin{equation}
\left|\widetilde{\phi}\right|_t \approx \left|\widetilde{\phi}\right|log_2~t/\tau. \label{eq_13}
\end{equation}
However there is a caveat here. When we consider that the time dependent magnetic field $\lambda (t<0)$ traverses the critical region $0 \leq \lambda \leq 2$ starting from $\lambda=0$ at $t=\tau$ and ending at $\lambda=2$ at $t=0$ in a closed circuit it is noted that a spin state while traversing this closed path will acquire the Berry phase factor $|\widetilde{\phi}|$. However when we transcribe the RG eqn.\eqref{eq_4} in terms of $t/\tau$ as shown in \eqref{eq_11} we note that $t/\tau$ is restricted in the region $0 \leq t/\tau \leq 1$. So when we map the closed circuit $0 \leq \lambda \leq 2$ onto the circuit $0 \leq t/\tau \leq 1$ the Berry phase factor acquired by a spin state while traversing this closed path will be half of that acquired in the former case. So for the effective phase factor we write
\begin{equation}
|\widetilde{\phi}|_{eff}=\frac{1}{2}|\widetilde{\phi}|, \label{eq_13a}
\end{equation}
and we have
\begin{eqnarray}
\Delta S &=& |\widetilde{\phi}|_{eff}~log_2~t/\tau, \nonumber \\
&=& \frac{1}{2}|\widetilde{\phi}|~log_2~t/\tau. \label{eq_13b}
\end{eqnarray}
Now from \eqref{eq_10} we have for the entropy variation with $\lambda$ from the value at $\lambda=0$
\begin{equation}
\Delta S(\lambda)=\frac{1}{2}|\widetilde{\phi}|log_2\left(1-\frac{\lambda}{2}\right). \label{eq_13c}
\end{equation}
As mentioned above the Berry phase factor $\widetilde{\phi}$ which corresponds to concurrence for a pair of nearest neighbor spins in the spin chain which at $\lambda=0$ corresponds to that in the antiferromagnetic system is given by $\widetilde{\phi}=0.386$. Incorporating the correction factor 0.926 we have the prefactor 0.35 which is very close to the value 1/3 derived from conformal field theory. So we write
\begin{eqnarray}
S_L(\lambda) &=& S_L(\lambda=0)+\frac{0.35}{2}~log_2\left(1-\frac{\lambda}{2}\right), \nonumber \\
& \approx & S_L(\lambda=0)+\frac{1}{6}~log_2\left(1-\frac{\lambda}{2}\right). \label{eq_15}
\end{eqnarray}
This is valid for all values of $L$ which satisfy the constraint \eqref{eq_9}. It may be noted that an analytical expression for the magnetic field dependence of the entanglement entropy has been given by Jin and Korepin~\cite{21}. It has been observed that their result is compatible with the numerical studies which fixes the value of the constant term to be added to the expression \eqref{eq_3}~\cite{22}.
\\
\noindent It is observed that the $\lambda$ dependent term in expression \eqref{eq_15} is independent of $\tau$ and thus will be valid for QPT without a quench. Indeed it is found to be in excellent agreement with the numerical results obtained by Lattore and Riera as shown in fig.2 in~\cite{22}. For QPT induced by a quench taking the expression for $S_L~(\lambda=0)$ as given by \eqref{eq_7} we can compute $S_L(\lambda)$.
\begin{figure}[htbp]
\centering
\includegraphics[height=10cm,width=15cm]{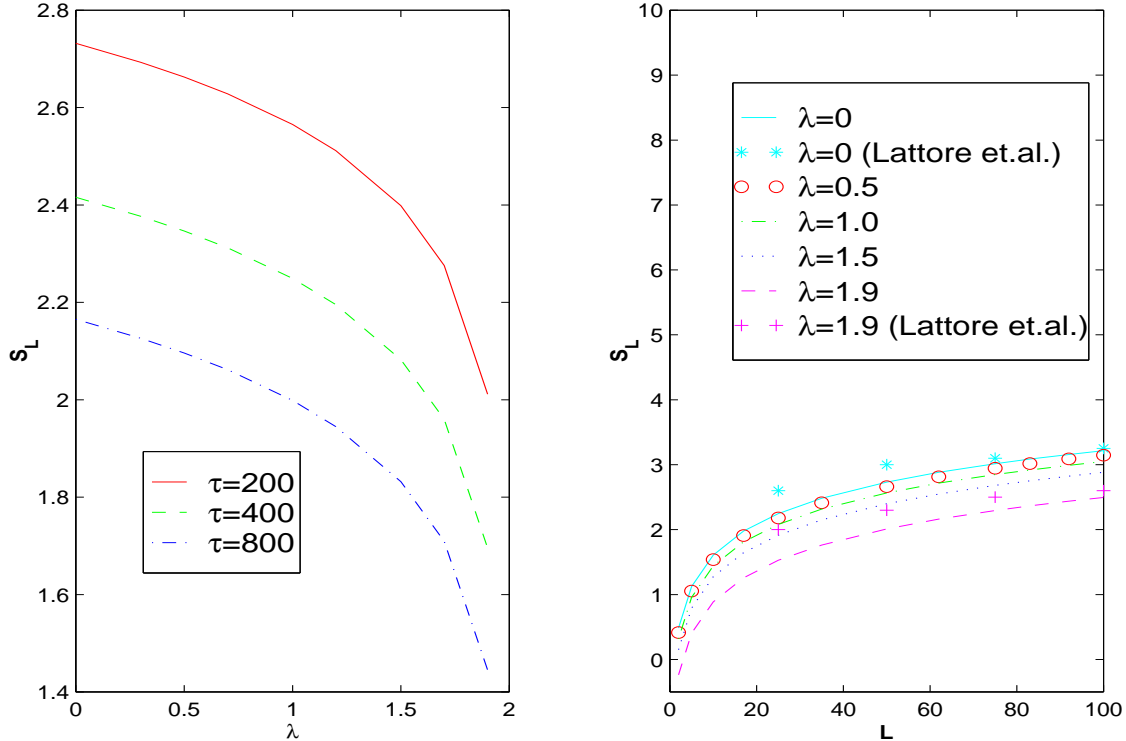}
\caption{\label{cap:fig_1} (Color online) Figure (left) shows $S_L$ vs. $\lambda$ for $L=50$ and $\tau=200,400,800$. Figure (right) depicts the variation of $S_L$ with $L$ for $\tau=200$ and $\lambda=0,0.5,1.0,1.5,1.9$. We have compared our results for $\lambda=0$ and $\lambda=1.9$ with that derived in \cite{22}.}
\label{fig_1}
\end{figure}
\\
\noindent In fig.\ref{fig_1} (left) we plot the variation of $S_L$ vs. $\lambda$ for a fixed value of $L=50$ and for different values of $\tau=200,400,800$. In fig.\ref{fig_1} (right) we plot $S_L$ vs. $L$ for a fixed value of $\tau=200$ and for different values of $\lambda=0,0.5,1.0,1.5,1.9$. We have compared our results with that obtained by Lattore and Riera~\cite{22} for $\lambda=0$ and $\lambda=1.9$ without any quench in the thermodynamic limit. From the fig.\ref{fig_1} (right) we note that for $\tau=200$ the results are found to be in very good agreement with their findings without the introduction of any quench. For higher values of $\tau$ the entropy decreases as follows from \eqref{eq_7}.
\\
\noindent The Hamiltonian for the XXX model is given by
\begin{equation}
H_{XXX}=\sum_i \left(\sigma_i^x\sigma_{i+1}^x+\sigma_i^y\sigma_{i+1}^y+\sigma_i^z\sigma_{i+1}^z\right)+\lambda \sum_i \sigma_i^z. \label{eq_16}
\end{equation}
The critical behavior of the model is analogous to that of the XX model. It has two-limit behavior. At $\lambda=2$ the system represents the ferromagnetic state and at $\lambda=0$ the system corresponds to the antiferromagnetic state. The interval $0<\lambda<2$ is gapless and hence critical. As at $\lambda=0$ this corresponds to the antiferromagnetic state the entanglement entropy for a block of $L$ spins with the rest of the system is identical with that of the XX model. This is valid for the quench induced QPT also. Just as the XX model here also the entanglement entropy is maximum at $\lambda=0$ and decreases with the increase in $\lambda$ until at $\lambda=2$ it vanishes. The variation of $S_L(\lambda)$ with different values of $\lambda$ will be identical with that of the XX model.

\vskip .5cm

\section{Entanglement entropy in the critical region of the Lipkin-Meshkov-Glick model:} \label{sec_3}

\begin{figure}[htbp]
\centering
\includegraphics[height=10cm,width=15cm]{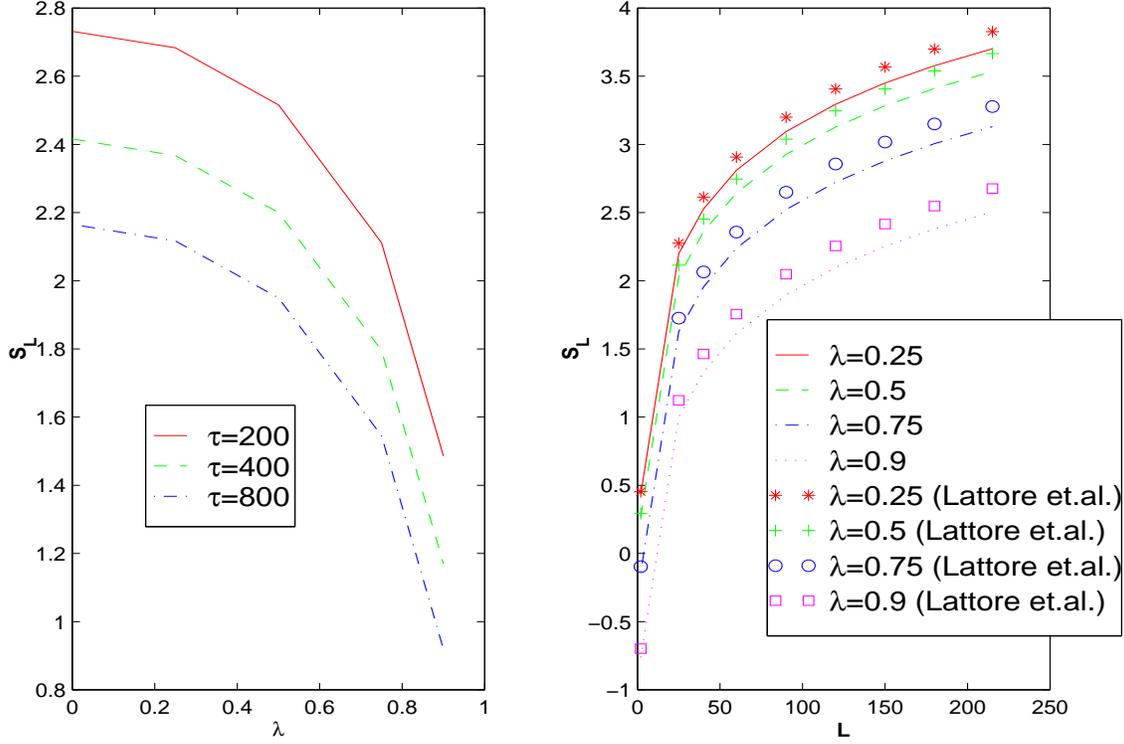}
\caption{\label{cap:fig_2} (Color online) Figure (left) depicts the variation $S_L(\lambda, \gamma=1)$ with $\lambda$ for a given $L=50$ and for different values of $\tau=200,400,800$. In fig.\ref{fig_2} (right) we have plotted the entropy vs. $L$ for a fixed value of $\tau=200$ and for different values of $\lambda=0.25,0.5,0.75,0.9$. The results are compared with that obtained in \cite{23} extrapolating it to the thermodynamic limit.}
\label{fig_2}
\end{figure}

\noindent The Hamiltonian for the Lipkin-Meshkov-Glick (LMG) model is given by~\cite{13}
\begin{equation}
H=\frac{1}{N}\sum_{i<j}\left(\sigma_i^x\sigma_{j}^x+\gamma \sigma_i^y\sigma_{j}^y\right)+\lambda \sum_{i} \sigma_i^z, \label{eq_17}
\end{equation}
$N$ being the total number of spins. In contrast to the conventional spin model in the LMG model each spin interacts with all the spins of the system with same coupling constant. This introduces the loss of the notion of geometry as there is no distance between the spins. Thus we cannot consider the notion of a block of spins as a set of contiguous spins here. The symmetry of the Hamiltonian suggests that the ground state belongs to a symmetric subspace where all the spins are indistinguishable and this subspace restricts the entanglement entropy of a block of $L$ spins with the remaining spins. However the scaling behavior shows a similar pattern with that of the XX model where conformal symmetry plays a significant role in the critical region. However in the LMG model though the scaling law is similar to that of the XX model it has nothing to do with the underlying conformal symmetry.
\\
\noindent In the isotropic case with $\gamma=1$ the Hamiltonian can be written in terms of the total spin operator $S^{\alpha}=1/2\sum \sigma_i^{\alpha}$ as
\begin{equation}
H=\frac{2}{N}\left(\vec{S}^2-(S^z)^2-\frac{N}{2}\right)+2\lambda S^z. \label{eq_18}
\end{equation}
Now as shown in an earlier paper~\cite{12} if we introduce point-splitting regularization so that we write
\begin{equation}
\vec{S}^2=\vec{S_k}.\vec{S_k'}\delta_{kk'}, \label{eq_19}
\end{equation}
where $k$ and $k'$ are two adjacent sites with an infinitesimal distance $k-k'=\epsilon $ and the relation \eqref{eq_19} is satisfied in the limit $\epsilon \rightarrow 0$, we can take
\begin{equation}
S_k^{\alpha}=\frac{1}{2}\sum \sigma_i^{\alpha},~ S_{k'}^{\alpha}=\frac{1}{2}\sum \sigma_j^{\alpha}, \label{eq_20}
\end{equation}
with $i$ and $j$ being two adjacent sites with an infinitesimal distance. Considering only nearest neighbor interactions we can take the Hamiltonian in the regularized form~\cite{12}
\begin{equation}
H_{reg}=\frac{1}{2N}\left[\sum_{i,j}\left(\sigma_i^x\sigma_{j}^x+\sigma_i^y\sigma_{j}^y+\sigma_i^z\sigma_{j}^z\right)+N\lambda \sum_{j} \sigma_i^z\right]-\frac{1}{2N}\left(\sum_{i,j}\sigma_i^z\sigma_{j}^z-N\lambda\sum_{j} \sigma_i^z\right)-1. \label{eq_21}
\end{equation}
From this we note that when we consider $1/2N$ as a dimensionless coupling constant $J$ we can write $H_{reg}=H_1+H_2-1$ with
\begin{eqnarray}
H_1 &=& J\left[\sum_{i,j}\left(\sigma_i^x\sigma_{j}^x+\sigma_i^y\sigma_{j}^y+\sigma_i^z\sigma_{j}^z\right)+\frac{\lambda}{2J}\sum_{j} \sigma_i^z\right], \label{eq_22} \\
H_2 &=& -J\left(\sum_{i,j}\sigma_i^z\sigma_{j}^z-\frac{\lambda}{2J}\sum_{j} \sigma_i^z\right) \label{eq_23}
\end{eqnarray}
The Hamiltonian given by \eqref{eq_21} can be split into various other $H_1$ and $H_2$ such that both of them describe critical systems. However as the LMG model is characterized by the fact that each spin interacts with every other spin so that for even (odd) number of interacting spins we have bosonic (fermionic) systems, we have chosen the regularized Hamiltonian such that one of them represent bosonic and the other fermionic features. From \eqref{eq_22} and \eqref{eq_23} we note that $H_1$ and $H_2$ represent the XXX model and Ising model respectively.
\\
\noindent To consider the criticality of the system we observe from \eqref{eq_21}, \eqref{eq_22} and \eqref{eq_23} that this is a combination of the XXX model given by $H_1$ and the Ising model with magnetic field along the negative $z$ direction given by $H_2$. It is observed that in this regularized form $\lambda/2J$ denotes the intensity of the magnetic field. The XXX model is characterized by the fact that for $2>|\lambda|/2J>0$ the system is gapless and hence critical. For $|\lambda|/2J=2$ we have $|\lambda|=2/N$ as we have $J=1/2N$. Since for an entangled spin state the minimum number of spins must be 2 we find that at criticality $|\lambda|$ lies in the interval $0<|\lambda|<1$. For the Ising chain given by $H_2$ the region $|\lambda|>1$ corresponds to the fact that all spins are polarized along the negative $z$-axis. In the interval $0<|\lambda|<1$ as $|\lambda|$ is tuned from 1 to 0 the spin system will undergo a transition when down spins will be excited so that at $\lambda=0$ all spins will settle down with opposite orientation. In between these two points with the tuning of $\lambda$ spins evolve through a paramagnetic state. Thus during the transition in the interval $0<|\lambda|<1$ spins evolve through a situation which is similar to that of the transverse Ising model. It is observed that when the LMG model is recast in the regularized form the fact that the critical region manifests the same logarithmic scaling law as observed in XXX model and the transverse Ising model can be understood from the conformal symmetry in the critical region in these systems. Indeed the point-splitting regularization unveils the underlying conformal symmetry at criticality in this system which is lost in the sharp point limit.
\\
\noindent In the regularized Hamiltonian introducing a quench we consider that the system transits from the point $\lambda=0$ where the entropy is maximum towards the value $|\lambda|=1$ where the entropy vanishes. So for QPT induced by a quench we consider that the time dependent magnetic field behaves as
\begin{equation}
\lambda(t<0)=1-\frac{t}{\tau}, \label{eq_24}
\end{equation}
so that at $t=\tau$, $\tau$ being the quench time, $\lambda=0$ and at the final state, when $t=0$ we have $\lambda=1$. Now from the RG flow \eqref{eq_4} and transcribing it in terms of $t/\tau$ we have for the variation of the entropy with $\lambda$ from the value at $\lambda=0$
\begin{equation}
\Delta S \approx |\widetilde{\phi}|log_2~t/\tau,   \label{eq_25}
\end{equation}
with $\widetilde{\phi}=|\widetilde{\phi}|_{XXX}+|\widetilde{\phi}|_{Ising}$. It has been shown in an earlier paper~\cite{11} that the Berry phase factor which is equivalent to the concurrence for the nearest neighbor spins in the transverse Ising model is given by $C=|\widetilde{\phi}|_{Ising}=0.18$. For the XXX model we have $|\widetilde{\phi}|_{XXX}=0.386$~\cite{14,15}. Thus $|\widetilde{\phi}|=0.386+0.18=0.566$. Taking into account the correction factor 0.926 as mentioned in the previous section we have the effective value of $|\widetilde{\phi}|_{eff}=0.52$. Now from \eqref{eq_24} we have $t/\tau=1-\lambda $. However we have to take care of the fact that in the regularized Hamiltonian for the XXX model the external field is in the positive $z$-direction while for the Ising model this is in the negative $z$-direction. So in the sharp point limit we have to take both these orientations of the magnetic field. Thus from \eqref{eq_25} we write in the sharp point limit for the entropy variation with $\lambda$ from the value at $\lambda=0$
\begin{eqnarray}
\Delta S(\lambda) &=& |\widetilde{\phi}|_{eff}\left[log_2(1-\lambda)+log_2(1+\lambda)\right], \nonumber \\
&=& 0.52~log_2(1-\lambda^2).   \label{eq_26}
\end{eqnarray}
Thus we have
\begin{equation}
S_L(\lambda,\gamma=1)=S_L(\lambda=0,\gamma=1)+0.52~log_2(1-\lambda^2), \label{eq_27}
\end{equation}
which is in excellent agreement with the value obtained numerically~\cite{23}
\begin{equation}
S_L(\lambda,\gamma=1)=S_L(\lambda=0,\gamma=1)+\frac{1}{2}~log_2(1-\lambda^2). \label{eq_28}
\end{equation}
It may be mentioned here that when QPT is induced by a quench the value of the entanglement entropy for the isotropic LMG model ($\gamma=1$) at $\lambda=0$ is found to be given by~\cite{12}
\begin{equation}
S_L(\tau) \approx 2\frac{\widetilde{\phi}log_2L}{\widetilde{\phi}log_2\widehat{\xi}}\times 0.926
\approx 3.7\frac{log_2L}{log_2\tau}.  \label{eq_29}
\end{equation}
From the constraint $S_L(\tau)/S_{max} \leq 1$ we find the constraint for $L$~\cite{12}
\begin{equation}
ln~L|_{LMG} \leq 0.097(ln~\tau)^2+0.5~ln~\tau. \label{eq_30}
\end{equation}
Within this restricted region of $L$ the entropy scales like $log_2L$. Thus within this restricted region of $L$ the RG flow will be valid. So from \eqref{eq_26} we can write explicitly
\begin{equation}
S_L(\lambda,\tau,\gamma=1)=S_L(\lambda=0,\tau,\gamma=1)+0.52~log_2(1-\lambda^2), \label{eq_31}
\end{equation}
with
\begin{equation}
S_L(\lambda=0,\tau,\gamma=1)\approx \frac{3.7~log_2L}{log_2\tau}, \label{eq_32}
\end{equation}
where $L$ is constrained by the relation \eqref{eq_30}.
\\
\noindent In fig.\ref{fig_2} (left) we plot the variation of $S_L$ with $\lambda$ for a fixed $L=50$ and for different values of $\tau=200,400,800$. In fig.\ref{fig_2} (right) we show the variation of $S_L$ with $L$ for a fixed $\tau=200$ and for different values of $\lambda=0.25,0.5,0.75,0.9$. The results are compared with that obtained in \cite{23} without introducing the quench extrapolating it to the thermodynamic limit. It is noted that with increase of $\tau$, the entropy decreases.
\\
\noindent In the anisotropic case ($\gamma \neq 1$) for the LMG model in the thermodynamic limit the ground state for $\lambda>1$ represents the product state. For $\lambda \rightarrow 0$ the entanglement entropy saturates and goes to a constant that depends on $\gamma$. For $\gamma=0$ the ground state is degenerate and lives in the subspace of $\pm x$ direction. These two different phases suggest the existence of a QPT between $\lambda>1$ and $\lambda<1$. We now note that $\gamma=0$ is the representative of the class $\gamma \neq 1$ and for $\gamma=0,~\lambda>0$ the regularized Hamiltonian corresponds to the transverse Ising model. Indeed from the relations \eqref{eq_17} and \eqref{eq_21} we note that for $\gamma=0$ there is an extra term $-J\sum \sigma_i^y\sigma_j^y$ in the second term on the r.h.s. of \eqref{eq_21}. This modifies the Hamiltonian $H_2$ in \eqref{eq_23} which is now given by the equation
\begin{equation}
\overline{H_2}=H_2(\gamma=0)=-J\left(\sum_{i,j}\sigma_i^z\sigma_j^z-\sigma_i^y\sigma_j^y-\frac{\lambda}{2J}\sigma_i^z\right). \label{eq_32a}
\end{equation}
The total regularized Hamiltonian
\begin{equation}
\overline{H}=H(\gamma=0)=H_1+\overline{H_2} \label{eq_32b}
\end{equation}
effectively corresponds to that of the transverse Ising model. Now from the relation \eqref{eq_24} we note that at $t=0$ it reaches the critical point $\lambda=1$ starting from $\lambda=0$ at $t=\tau$ and we find the entropy around criticality in the thermodynamic limit
\begin{equation}
S_L (\lambda,\gamma \neq 1) \approx \frac{1}{6}log_2(1-\lambda). \label{eq_33}
\end{equation}
The prefactor 1/6 corresponds to the prefactor associated with the entanglement entropy of the transverse Ising model~\cite{11}. This is identical with the result obtained by Lattore et.al.~\cite{23} from numerical studies. It is to be mentioned that just like the isotropic case for QPT induced by a quench, $L$ is restricted by the constraint given by \eqref{eq_30}.

\vskip .5cm

\section{Variation of entropy with the anisotropy parameter:} \label{sec_4}

\noindent We extend our study here when QPT is subject to a quench across quantum multicritical points by approaching along a linear path formulating the anisotropy parameter dependence of the external magnetic field~\cite{24}. We take
\begin{equation}
\lambda(\gamma)=1-\gamma(t),~~t<0 ~~~(\gamma \neq 1). \label{eq_34}
\end{equation}
The XY model Hamiltonian is given by
\begin{equation}
H_{XY}=-\sum_i\left(\frac{1+\gamma}{2}\sigma_i^x\sigma_{i+1}^x+\frac{1-\gamma}{2}\sigma_i^y\sigma_{i+1}^y+\lambda \sigma_i^z\right). \label{eq_35}
\end{equation}
For $\gamma=1$ it corresponds to the transverse Ising model. For $\gamma \neq 0$ the system falls into the free fermion universality class and is critical at $\lambda=1$. For $\gamma=0$ the system reduces to the XX system which at $\lambda=0$ corresponds to the free boson class. To study the $\gamma$ dependence of the entropy we take into account the expression \eqref{eq_34} exhibiting the $\gamma$ dependence of $\lambda$ implying multicritical behavior. When we introduce a quench we take that at $t=\tau$, $\tau$ being the quench time the critical point $\gamma=1$ is reached so that we have
\begin{equation}
\gamma(t)=-\frac{t}{\tau},~~~t<0. \label{eq_36}
\end{equation}
From \eqref{eq_34}, this implies
\begin{equation}
\lambda(t<0)=1-\frac{t}{\tau}, \label{eq_37}
\end{equation}
so that at $t=\tau$ we have $\lambda=0$. The critical point $\lambda=1$ is reached for the fermion universality class at $t=0$ by evolving from the base point $\lambda=0$ at $t=\tau$. At the critical point the entanglement entropy for the transverse Ising model follows the scaling law~\cite{1}
\begin{equation}
S_L \sim \frac{1}{6} log_2L, \label{eq_38}
\end{equation}
which follows from the relevant conformal field theory. From the relationship between the Berry phase factor obtained by a spin state when it evolves in a closed path with the concurrence for a pair of nearest neighbor spins the prefactor is found to be 0.18~\cite{7} which is very close to the factor 1/6. Indeed this determines the correction factor $(1/6)/0.18=0.926$ introduced earlier which incorporates the correction for block variable RG scheme. Now as discussed in sec.\ref{sec_2} from the RG flow equation \eqref{eq_4} and transcribing it in terms of $t/\tau$ we have from \eqref{eq_36} the variation of the entropy $S$ with $\gamma$ from the value of the entropy at $\lambda=1,~\gamma=1$
\begin{eqnarray}
\Delta S(\gamma) &=& \frac{1}{6}log_2~t/\tau, \nonumber \\
&=& \frac{1}{6}log_2~\gamma. \label{eq_39}
\end{eqnarray}
Thus we write
\begin{equation}
S_L(\lambda=1,~\gamma)=S_L(\lambda=1,~\gamma=1)+\frac{1}{6}log_2\gamma, \label{eq_40}
\end{equation}
which is satisfied for the fermion universality class ($\gamma \neq 0$) in the thermodynamic limit. This result is identical with that obtained by Vidal, Lattore, Rico and Kitaev~\cite{1} from a detailed computational study. It is observed that the singular behavior at $\gamma=0$ is due to the fact that at this point the system represents the XX model which at $\lambda=0$ corresponds to the free boson class.
\\
\noindent It has been shown in an earlier paper~\cite{11} that when QPT is induced by a quench the scaling law of the entanglement entropy of the transverse Ising model ($\gamma=1$) at the critical region is given by
\begin{equation}
S_L(\tau)\approx \frac{3.7log_2L}{log_2 \tau}, \label{eq_41}
\end{equation}
and the validity of this scaling property is restricted to the region given by the condition $S_L(\tau)/S_{max} \leq 1$ which implies
\begin{equation}
ln~L \leq 0.03(ln~\tau)^2+0.5~ln~\tau. \label{eq_42}
\end{equation}
Within this restricted region of $L$ the RG flow equation is valid and from \eqref{eq_40} we can write for $\gamma \neq 0$
\begin{equation}
S_L(\gamma)\approx \frac{3.7log_2L}{log_2 \tau}+\frac{1}{6}log_{2}\gamma. \label{eq_43}
\end{equation}
\begin{figure}[htbp]
\centering
\includegraphics[height=10cm,width=15cm]{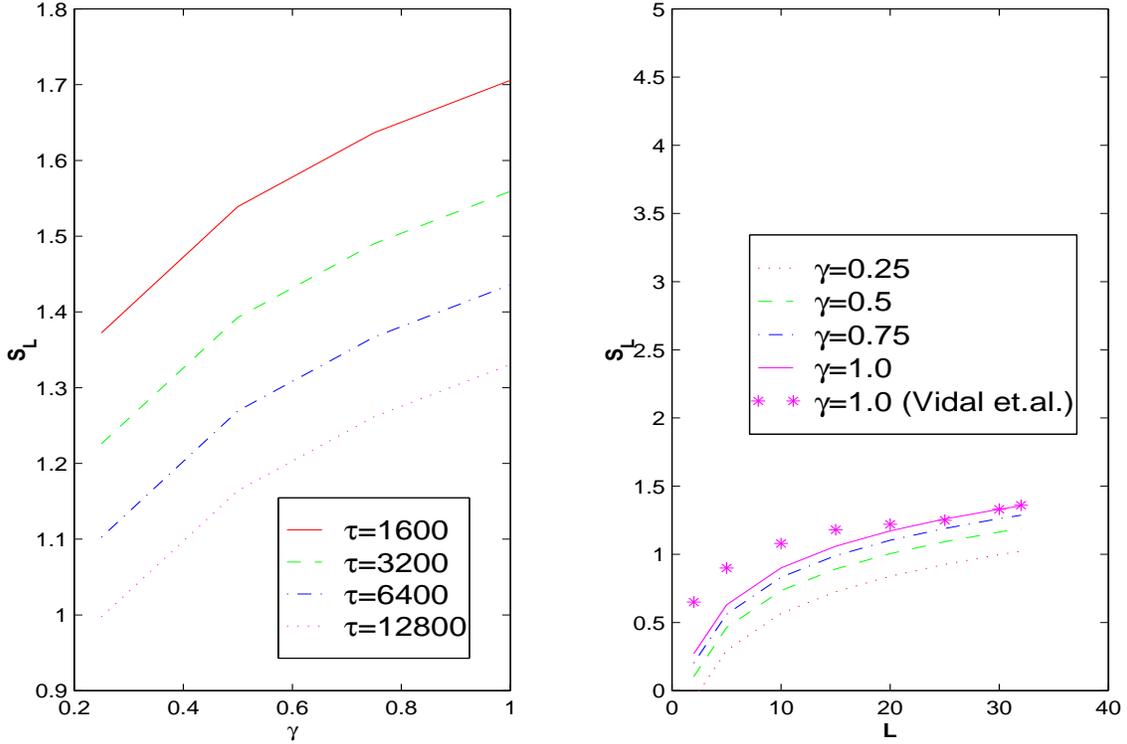}
\caption{\label{cap:fig_3} (Color online) Fig.3 (left) shows the variation $S_L$ with $\gamma$ for a fixed value of $L=30$ and for different values of $\tau=1600,3200,6400,12800$. Fig.\ref{fig_3} (right) shows the variation of $S_L$ with $L$ for a fixed $\tau=12800$ and for different values of $\gamma=0.25,0.5,0.75,1$. The results are compared with that obtained in \cite{1} for $\gamma=1$.}
\label{fig_3}
\end{figure}
To compare it with the exact value of the entropy obtained numerically by Vidal et.al.~\cite{1} at $\gamma=1$ in QPT without quench it may be mentioned that the studies in \cite{1} involve infinite Ising chain. However the relation \eqref{eq_42} implies that the maximum value of $L$ is restricted by the quench time $\tau$ and indicates that for small $\tau$, $L_{max}$ is very small. In view of this we take large $\tau$ for comparison. In fig.\ref{fig_3} (left) we plot the entanglement entropy vs. $\gamma$ for a fixed $L=30$ and different values of $\tau=1600,3200,6400,12800$. In fig.\ref{fig_3} (right) we show the entanglement entropy vs. $L$ for a fixed $\tau=12800$ and for different values of $\gamma=0.25,0.5,0.75,1$. It is observed that for $\gamma=1$ our results are in good agreement with that of Vidal et. al. \cite{1} derived from numerical studies without introducing any quench. As in other spin systems we note that the entropy decreases with the increase in $\tau$.
\\
\noindent It may be mentioned here that a specific property of QPT in transverse Ising model is that the maximum of the concurrence for the entanglement of a pair of nearest neighbor spins does not occur at the critical point $\lambda=1$ but away from it. Indeed at the critical point the correlation length diverges and each site develops entanglement with its nearest neighbor site. In view of this the entanglement of a certain pair of nearest neighbor sites is distributed in such a way that it saturates the constraints of entanglement sharing. Osborne and Nielsen~\cite{5} have conjectured that the ground state at the critical point actually saturates the bounds of entanglement sharing so that it is maximally entangled in this sense. This would imply that when the system reaches criticality the entanglement is distributed to more remote pairs and entanglement sharing would have to occur at the expense of the two party entanglement. In our present analysis we note that near criticality for the XY model with $\gamma \neq 0$ the variation of the entanglement entropy with $\lambda$ is given by
\begin{equation}
S=\frac{1}{6}\left|log_2(1-\lambda)\right|  \label{eq_43a}
\end{equation}
which follows from \eqref{eq_37} and \eqref{eq_39}. From this we observe that near criticality the concurrence for a pair of nearest neighbor spins satisfies the relation
\begin{equation}
C=\frac{1}{12}\left|log_2(1-\lambda)\right|  \label{eq_43b}
\end{equation}
Here the prefactor is taken to be half of that in \eqref{eq_43a} due to the fact that a spin can form a pair with its nearest neighbor in both left and right sides and the effect of this is incorporated in the expression of the entanglement entropy~\cite{7,12}. Thus we note that we have a logarithmic divergence at the critical point $\lambda=1$. This indicates that at this point the correlation length diverges and entanglement is shared among the remote sites. As the concurrence decreases along the RG flow the maximum will occur at a point $\lambda<1$ close to the critical point $\lambda=1$.
\\
\noindent In a similar way we can analyze the LMG model given by the Hamiltonian  \eqref{eq_17}. For $\lambda=1,~\gamma \neq 1$ we can derive the variation of the entropy with $\gamma$ from the value at $\lambda=1,~\gamma=0$ using the fact that $\gamma=0$ represents the class $\gamma \neq 1$ and $\lambda=1$ is the critical point. For the evolution of $\gamma$ with time we take the relation $\gamma=1-t/\tau$ which implies that at $t=\tau$ we have the base point $\gamma=0$ and it evolves with time so that at the end $t=0$ it reaches the point $\gamma=1$. From the RG flow equation \eqref{eq_4} the deviation of entropy for $\gamma \neq 1$ from the value at $\lambda=1,~\gamma=0$ is found to be
\begin{eqnarray}
\Delta S(\lambda=1,\gamma) &=& \frac{1}{6}log_2~t/\tau, \nonumber \\
&=& \frac{1}{6}log_2(1-\gamma), \label{eq_44}
\end{eqnarray}
so that we have
\begin{equation}
S_L(\lambda=1,\gamma)=S_L(\lambda=1,\gamma=0)+\frac{1}{6}log_2(1-\gamma). \label{eq_45}
\end{equation}
This is valid for all $\gamma$ in the region $-1 \leq \gamma < 1$. The prefactor 1/6 follows from the fact that at $\lambda=1,~\gamma=0$ the regularized Hamiltonian corresponds to the critical transverse Ising model where we have the prefactor 1/6 for the scaling law. This is consistent with the result obtained by Lattore et. al.~\cite{23} from numerical studies. Now to compute the entanglement entropy for the LMG model at $\gamma=0$ in QPT induced by a quench, we note that the prefactor depends on the quench time which arises from the fact that the final state represents a kink-antikink chain with lattice spacing approximately given by the Kibble-Zurek correlation length $\widehat{\xi}$. However in the LMG model a correlation length characterizing the typical distance between defects cannot be introduced though we can estimate the fraction of flipped spins after the quench. Now from the regularization scheme for $\gamma=0$ we note that the regularized Hamiltonians for $\gamma=1$ and $\overline{H}$ for $\gamma=0$ differ by a term $-J\sum_{i,j}\sigma_i^y\sigma_j^y$ as is evident from \eqref{eq_23} and \eqref{eq_32a}. This essentially introduces for the contribution of $\overline{H}_2$ in \eqref{eq_32a} an extra component corresponding to the transverse Ising model. Evidently the value of $|\widetilde{\phi}|_{eff}$ in the prefactor $|\widetilde{\phi}|log_2\widehat{\xi}$ associated with the quench time in the denominator of \eqref{eq_29} will be given by
\begin{equation}
|\widetilde{\phi}|_{eff}=|\widetilde{\phi}|_{XXX}+|\widetilde{\phi}|_{Ising}=(0.358+0.18)\times 0.926=0.52. \label{eq_46}
\end{equation}
where 0.926 is the correction factor as introduced above. This is essentially identical with the corresponding expression for LMG model with $\gamma=1$ as in \eqref{eq_29}. This term arises from the fraction of defects formed during critical slowing down and the regularization scheme suggests that this remains unaltered irrespective of the value of $\gamma$. However as for $\gamma=0$, the total regularized Hamiltonian essentially corresponds to that of the transverse Ising model the scaling law of the entanglement entropy at $\lambda=1,~\gamma=0$ will be identical with that model viz. $1/6log_2L$. This implies that the numerator in \eqref{eq_29} will be modified now and is given by $1/6log_2L$. Thus for QPT induced by a quench, the entanglement entropy for the LMG model at $\gamma=0$ can be expressed as
\begin{eqnarray}
S_L(\lambda=1,\gamma=0)|_{LMG} & \approx & 2\frac{\frac{1}{6}log_2L}{0.52~log_2\widehat{\xi}} \times 0.926 \nonumber \\
& \approx & 3.7\frac{\frac{1}{6}log_2L}{0.52~log_2\tau} \label{eq_47}
\end{eqnarray}
This suggests that we have the relation
\begin{equation}
S_L(\lambda=1,\gamma \neq 1)|_{LMG} \approx 3.7\frac{\frac{1}{6}log_2L}{0.52~log_2\tau}+\frac{1}{6}log_2(1-\gamma) \label{eq_48}
\end{equation}
as follows from \eqref{eq_45}. It is to be mentioned that $L$ is here restricted by the constraint \eqref{eq_30}.
\begin{figure}[htbp]
\centering
\includegraphics[height=10cm,width=15cm]{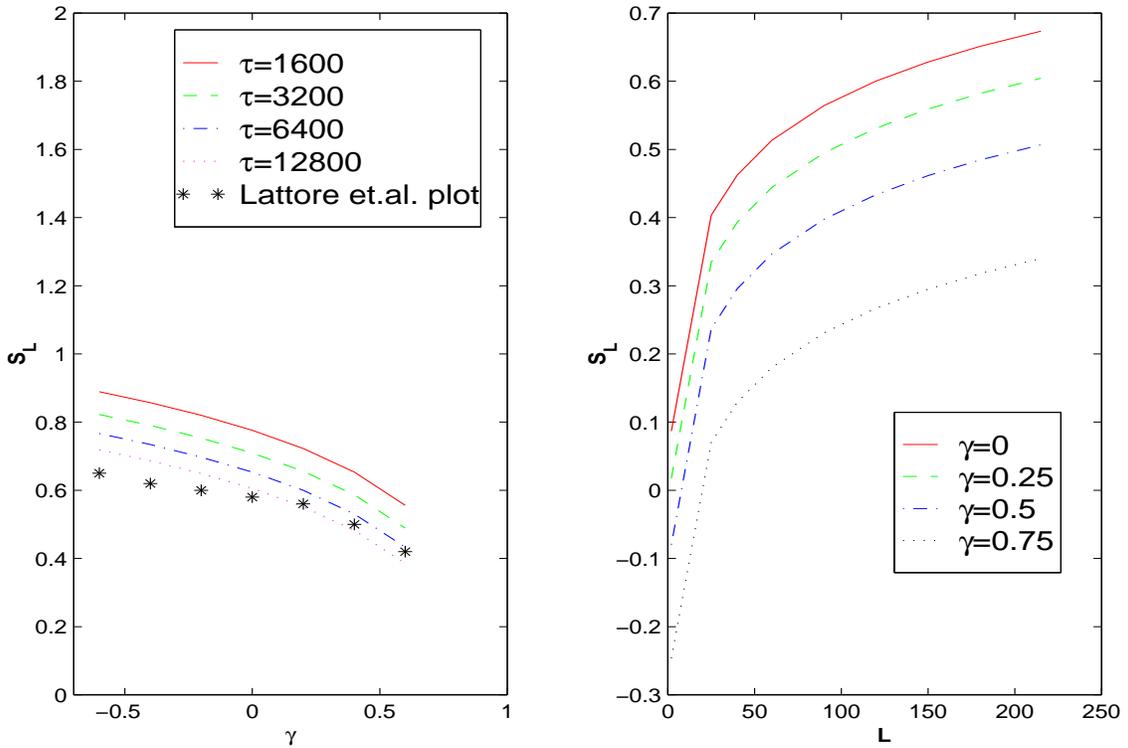}
\caption{\label{cap:fig_4} (Color online) Fig.4 (left) shows the variation of the entanglement entropy with $\gamma$ for $L=125$ and $\tau=1600,3200,6400,12800$. Fig.\ref{fig_4} (right) shows the variation of the entropy with $L$ for $\tau=12800$ and  $\gamma=0,0.25,0.5,0.75$. In fig.\ref{fig_4} (left) we have compared our results with that of Lattore et. al.~\cite{23} derived from numerical studies without introducing quench extrapolating it to the thermodynamic limit.}
\label{fig_4}
\end{figure}
\\
\noindent In fig.\ref{fig_4} (left) we show the variation of the entanglement entropy with $\gamma$ for a fixed value of $L=125$ and different values of $\tau=1600,3200,6400,12800$. In fig.\ref{fig_4} (right) we plot the variation of the entropy with $L$ for a fixed value of $\tau=12800$ and different values of $\gamma=0,0.25,0.5,0.75$. In fig.\ref{fig_4} (left) we have compared our results for $\tau=12800$ with that of Lattore et.al.~\cite{23} and is found to be in good agreement with those results derived numerically without introducing quench. As it has been pointed out that for $\gamma \neq 1$ the model represents the class for $\gamma=0$ and for $\lambda \neq 0$ the regularized Hamiltonian corresponds to that of the transverse Ising model, this explains the fact that the LMG model for $\gamma \neq 1$ as $\lambda$ varies away from the critical value is analogous to the transverse Ising model as observed by Lattore et. al. \cite{23}.

\vskip .5cm

\section{Discussion:}  \label{sec_5}

\noindent It may be mentioned that the logarithmic scaling law was derived for geometric entropy for a conformal field theory~\cite{2} and this expression has been confirmed for several critical spin chains~\cite{1}. The present analysis suggests that the logarithmic scaling law of the entanglement entropy of 1D spin systems is a consequence of the RG flow which follows from \eqref{eq_4}-\eqref{eq_6}. However there are situations which do not obey the logarithmic scaling law. Indeed Calabrese et.al.~\cite{25,26} have pointed out that after a global sudden quench the entanglement entropy increases linearly with time. These authors have shown that when the system is prepared in a pure state $\left|\psi_0\right\rangle$ which corresponds to an eigenstate of the Hamiltonian $H(\lambda_0)$ with $\lambda_0 \neq \lambda$ and at time $t=0$ the parameter is suddenly quenched from $\lambda_0$ to $\lambda$, then the entanglement entropy of 1D spin systems in an interval of length $l$ increases linearly with time up to a certain point after which it saturates at a value proportional to $l$ with a coefficient depending on the initial state. The behavior has been interpreted as a consequence of causality. In this context it may be added that for a quench induced QPT the logarithmic scaling behavior persists, though in a restricted sense, where there is a critical slowing down and the control parameter $\lambda$ is a linear function of time. In fact within the scaling region, we have the RG flow of the entropy. However for a sudden quench the RG equation will not be satisfied and hence we have a deviation from the logarithmic scaling law.
\\
\noindent In some earlier papers~\cite{11,12} it has been pointed out that when QPT in one dimensional spin system is induced by a quench the entanglement entropy satisfies a scaling law with a prefactor which depends on the quench time. However in this case for the entropy of a block of $L$ spins with the rest of the system the block size $L$ is restricted by a constraint. From the RG flow equation we have derived here the variation of the entropy with external magnetic field. The results are found to be in excellent agreement with that obtained from numerical studies by other authors without the introduction of the quench. For the LMG model we have taken into account the point splitting regularization and we have deduced our result from the regularized Hamiltonian. This unveils the underlying conformal symmetry at criticality which is lost at the sharp point limit.
\\
\noindent For the XY model and the LMG model we have studied the behavior of the entanglement entropy with the variation of the anisotropy parameter $\gamma$ when QPT is subject to a quench across quantum multicritical points by approaching along a linear path formulating the anisotropy parameter dependence of the external magnetic field. Using the RG flow equation we have considered the anisotropy parameter dependence of the entropy and is found to be in good agreement with that obtained by other authors~\cite{1,23}. For the LMG model as $\gamma=0$ is representative of the class $\gamma \neq 1$ and for $\gamma=0$ the regularized Hamiltonian corresponds to that of the transverse Ising model, the surprising behavior that for $\gamma \neq 1$ and as the magnetic field parameter $\lambda$ departs from the critical value the system is analogous to the transverse Ising model is well explained.
\\
\noindent Finally we point out that the RG flow equation for the central charge as well as for the Berry phase factor which is attained by a spin state when it evolves in a closed path essentially determines the RG flow equation of the entanglement entropy in QPT. This appears to be a strong tool to study the magnetic field dependence as well as the anisotropy parameter dependence of the entanglement entropy.

\end{document}